\documentclass[12pt]{iopart}
\usepackage{iopams}
\usepackage{graphicx}

\begin{document}

\title{Multiple timescales in a model for DNA denaturation dynamics}

\author{Marco Baiesi$^{1,2}$ and Roberto Livi$^2$}

\address{$^1$ Instituut voor Theoretische Fysica, K.~U.~Leuven, 
B-3001, Belgium.}
\address{$^2$ Dipartimento di Fisica, Universit\`a di Firenze,
and Sezione INFN, Firenze, I-50019 Sesto Fiorentino, Italy.}

\date{\today}

\begin{abstract}
The denaturation dynamics of a long double-stranded DNA is studied by means of
a model of the Poland-Scheraga type.
We note that the linking of the two strands is a locally conserved quantity,
hence we introduce local updates that respect this symmetry.
Linking dissipation via untwist is allowed only at the two ends of the 
double strand.
The result is a slow denaturation characterized by two time scales
that depend on the chain length $L$.
In a regime up to a first characteristic time $\tau_1\sim L^{2.15}$
the chain embodies an increasing number of small bubbles.
Then, in a second regime, bubbles coalesce and form entropic barriers that 
effectively trap residual double-stranded segments within the chain, 
slowing down the relaxation to fully molten configurations,
which takes place at $\tau_2\sim L^3$.
This scenario is different from the picture in which the helical
constraints are neglected.
\end{abstract}


\maketitle

The Watson-Crick helix is the typical form of DNA in the 
cell~\cite{Alberts02:_book}.
In the laboratory, upon heating DNA molecules in solution, 
one obtains an helix-coil transition called denaturation.
Since decades, DNA denaturation has attracted the 
attention of scientists because it can help to understand
important biological processes:
for instance,
the genetic code can be accessed during transcription and replication
by an opening of bubbles~\cite{Alberts02:_book}. 
It is experimentally known that the
fraction of molten DNA increases for increasing 
temperature~\cite{Wartell85}.
The first theoretical description of denaturation that can account for this
phenomenon was a simple model by Poland and Scheraga (PS)~\cite{PS_JCP1966}.
A model of this kind is now behind software like 
Meltsim~\cite{Blake99:_meltsim}, 
predicting sequence-dependent melting curves, which can then be compared with
experiments.
Another simple model for DNA is due to Peyrard and Bishop 
(PB)~\cite{PB_PRL1989,Barbi_PRE2003}.
Also some more detailed yet mesoscopic models have been recently
proposed~\cite{Drukker_WS_JCP2001,Knotts_JCP2007},
allowing for the numerical study of features that cannot be simulated in
all-atoms molecular dynamics. 

PS models rarely take into account the topological 
state of a macromolecule. All polymers that form closed rings have some
conserved topological features, as long as the chains cannot break
and cross each other. For instance,
in circular DNA, such as the genomes of some bacteria, 
the number of times that the two strands twist around each other 
(linking number) cannot change. 
If this constraint is included in a PS model,
the thermodynamic denaturation transition is essentially 
suppressed~\cite{RudnickBruinsma_PRE2002}, unless supercoiling effects
are considered~\cite{alkan2008}.
The topology is not fundamental for the equilibrium properties of
linear polymers,
but the fact that chains cannot cross each other is clearly relevant 
for their dynamics. For instance, in electrophoresis the dynamics
of DNA under an electric field depends on its continuous 
entangling with the polymers of a gel~\cite{Heukelum02:_electroph}. 
This feature is taken into account in models of 
electrophoresis~\cite{Heukelum02:_electroph}, like the Rubinstein-Duke 
``repton'' model~\cite{Rubinstein87,Duke89,Carlon01:_RDmodel}.

While the equilibrium thermodynamics of DNA has been largely investigated, 
the dynamics of this macromolecule is of interest as well. 
The dynamics of thermal denaturation has been studied
by means of the PS and similar 
models~\cite{Marenduzzo_2002, Kunz_LS_2007,Hanke03:_bubble_dyn,
Novotny06:_vicious}, 
by using the PB model~\cite{Barbi_PRE2003}, and in more detailed 
models~\cite{Drukker_WS_JCP2001,Knotts_JCP2007}.
For short DNA segments, the rates of opening 
found in experiments~\cite{Altan-bonnet03} can be estimated by
PS models with stochastic dynamics~\cite{Hanke03:_bubble_dyn}.
However, being a coarse grained description, the PS model is 
particularly useful to study long chains.

The aim of this paper is the study of the denaturation dynamics of a long
DNA in a PS model with a stochastic evolution that takes into account 
the helical structure of the double strand.
This was not explicitly included in the standard formulation of the PS 
model~\cite{Marenduzzo_2002, Kunz_LS_2007}.
We show that a dynamics with local preservation of the linking between
the chains yields a new scenario, with two time scales. 
A first regime is dominated by denatured bubbles diffusing into
the chain from its ends (where the double chain can freely untwist). 
This regime ends after a time lapse that scales approximately
as the square of the chain 
length, where the number of bubbles reaches a maximum. 
Then bubbles start to coalesce, with the ones near the chain ends that form 
entropic barriers trapping the helical domains still in excess.
The trapping  into these metastable states further slows down the 
denaturation process in the model, 
leading to a thermally equilibrated denaturation only after a second 
time scale, which grows as the cube of the chain length.

\begin{figure}[!tb] 
\vskip 1truemm
\begin{center}
\includegraphics[angle=0,width=8cm]{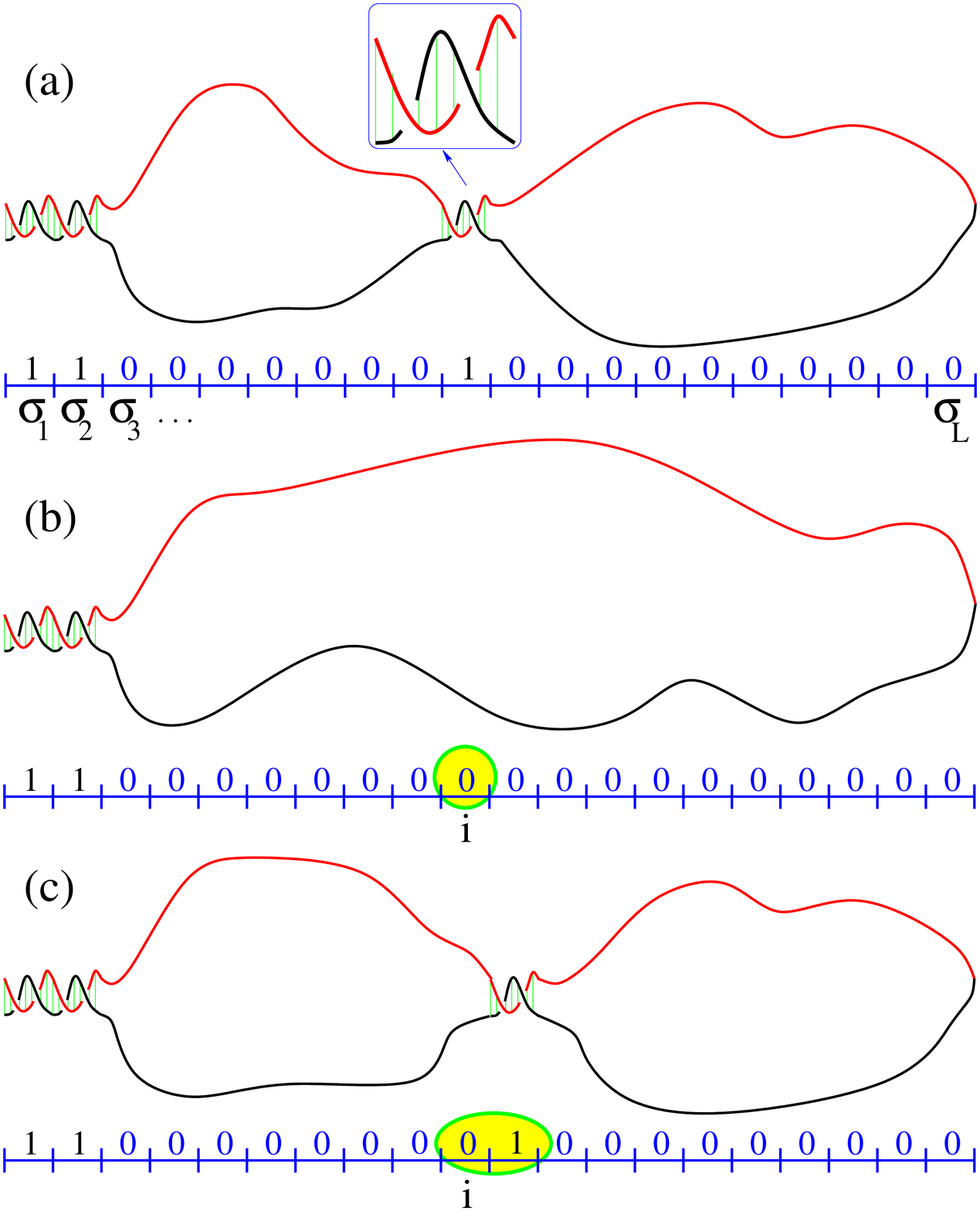}
\end{center}
\caption{(Color online)
(a) Sketch of a DNA configuration and of the relative array $\sigma$.
(b) Configuration obtained by updating (a)  
with (\ref{eq:1to0}) at site $i$.
(c) Configuration obtained by updating (a) 
with (\ref{eq:10to01}) at the same site: in this case there is a local
conservation of the linking number.
\label{fig:1}} 
\end{figure}

In the PS model, each strand is represented by a chain of length $L$, 
where a site $i$ ($1\le i \le L)$ stands for a local portion of DNA. 
We associate each site to a segment of 10 base
pairs, which thus represents a complete helical turn when 
paired~\footnote{Note 
that this choice of the coarse graining is not fundamental, 
as we are just interested in
the scaling properties of long chains and not in the  microscopic details
of the dynamics.}.
The state of the chain is stored in a Boolean array $\sigma_i$,
 where $\sigma_i=1$ if the two segments at index $i$ are paired and
$\sigma_i=0$ otherwise [see Fig.~\ref{fig:1}(a)]. 
DNA conformation are thus represented by an alternation of segments of paired 
bases (sequences of $1$'s)  and of open bubbles (sequences of $0$'s).
For every $\sigma_i=1$ there is a binding energy $\epsilon =-1$.
At a temperature $T$ this corresponds to a Boltzmann factor
$q=e^{-\epsilon/T}$.
Thus, a sequence of $m$ paired bases brings a contribution $q^m$ 
to the global weight $W$ of the configuration.
A bubble formed by two complementary strands of length $\ell$ 
instead has an entropic contribution accounting for all the
possible conformations of a walk of length $2\ell$: if $s$ is the entropy per 
step of a walk, the constraint to form a loop yields a weight 
$B s^{2\ell}\ell^{-c}$, where $B$ is a constant factor.
The exponent $c$ can be deduced from self-avoiding walks statistics:
with the excluded volume between the chains fully taken into 
account~\cite{Kafri_MP_2000} it is $c\approx 2.1$.
Hence, the weight of a whole configuration is 
\begin{equation}
W = (q^{m_1}) (B s^{2\ell_1}  \ell_1^{-c}) 
\ldots  
    (q^{m_\nu}) (B s^{2\ell_\nu}\ell_\nu^{-c}) (q^{m_{\nu+1}})
\label{eq:weight}
\end{equation}
where $\nu$ is the number of bubbles.
We also set $\sigma_0 = \sigma_{L+1} = 1$, namely
each end of a single strand is joined to the corresponding end of the other 
strand (no Y-fork is formed). Thus, at high $T$ the equilibrium 
configuration is a ring of length $2 (L+1)$.
At low $T$, on the other hand, typically one finds long double-stranded parts
separated by small bubbles.
According to this description, the properties of the 
model at thermodynamic equilibrium can be derived 
analytically~\cite{PS_JCP1966,Kafri_MP_2000}.

The simplest dynamical rules that can be assigned to the PS model involve moves
where locally one $\sigma_i$ changes,
\begin{equation}
\sigma_i=1 \qquad \longleftrightarrow \qquad \sigma_i=0 \;.
\label{eq:1to0}
\end{equation} 
A Metropolis criterion can then be used to choose whether to accept the move.
This kind of update  resembles the dynamics
of adsorption of a polymer onto a wall, see Fig.~\ref{fig:1}(a)-(b).
However, for $1\ll i\ll L$,
we note that the (dis)appearance of a ``1'' 
would imply either a temporary breaking of one of the chains to 
(un)twist the two strands there (as it happens e.g.~with 
topoisomerase enzymes~\cite{Dean85:_topois}) or
a global rotation  of $2\pi$ of the whole part $<i$ of the chain
with respect to the whole part $>i$. The latter case is not in agreement
with the idea of small time step that is intrinsic in (\ref{eq:1to0}).
Since the update (\ref{eq:1to0}) neglects the helix 
(and the consequent link) of the DNA strands, one needs another dynamics
that respects the local topology of dsDNA.

In order to preserve locally the linking number, we
adopt a different basic move: one picks a boundary 
$(i|i+1)$ at random and swaps the relative variables,
\begin{equation}
\sigma_i=x,\,\sigma_{i+1}=y
\quad \longrightarrow \quad 
\sigma_i=y,\,\sigma_{i+1}=x \;,
\label{eq:10to01}
\end{equation} 
where $x$ and $y$ can be $0$ or $1$ 
(if they are equal, the move is trivially the identity).
This exchanges the amount of linking of the chains at position $i$
with that at position $i+1$, see the sketch in Fig.~\ref{fig:1}(c).
In practice, the linking plays here the role of an order parameter that is
locally conserved. In order to have an evolution of the global linking number,
however, one has to consider also some dissipation via untwisting of the
double helix. 
In fact, untwisting with an update like (\ref{eq:1to0}) 
should be valid close to the ends of the dsDNA macromolecule, 
if they are free to rotate.
Two special updates are then introduced by adding to the list of possible
boundaries the $(0|1)$ and the $(L|L+1)$ ones:
\begin{eqnarray}
\sigma_1 \to 1-\sigma_1 && \qquad \textrm{if $(0|1)$ is chosen}\label{eq:s1}\\
\sigma_L \to 1-\sigma_L && \qquad \textrm{if $(L|L+1)$ is chosen}\label{eq:sL}
\end{eqnarray}
These two moves can thus change the energy 
$E = \sum_i \sigma_i$ of the chain,
allowing equilibration at every temperature. 
It is important to note that the move (\ref{eq:10to01}) 
can lead to the nucleation of new bubbles along the whole chain, from the 
boundaries of already present ones (e.g. $\sigma=\ldots00001111\ldots$ 
$\to$ $\sigma=\ldots00010111\ldots$). 
The dynamics obeys detailed balance, hence
clearly also the reverse process of bubble coalescence can take place.
Therefore,  move (\ref{eq:10to01}) seems the best approximation for
the purpose of describing the local conservation of the linking between two
complementary DNA chains. 
By definition this model cannot deal with the formation of twisted bubbles,
for example by means of a breaking of base-pair bonds without untwisting
the chain. These configurations, however, are entropically unfavored compared
to the ones in which the twist is concentrated on paired segments and
open bubbles are expanded. We thus expect that the approximation of neglecting
the formation of twisted bubbles is appropriate in a simple model.

A time step consists in a sequence of $L$ realizations of a basic move,
each one with $i$ picked at random, uniformly along the chain.
If the dynamics (\ref{eq:10to01})-(\ref{eq:sL}) is used,  
$i\in [0,L]$, while $i\in [1,L]$ for update (\ref{eq:1to0}).
Then, according to the Metropolis criterion each move is
accepted with probability $p = \min\{1, W_{\rm{new}} / W_{\rm{old}}\}$, 
where $W_{\rm{new}}$
is the weight of the proposed configuration and $W_{\rm{old}}$ 
is the weight of the present configuration.

The protocol on which we focus is a quench of a system
equilibrated at low $T$ to a regime at very high $T$.
This is indeed the regime that allows us to appreciate more the 
effects of the new physical ingredients in this model. 
Equilibrated configurations to start the protocol are generated 
by setting $q/s^2 = 100$ and by applying multiple (\ref{eq:1to0}) 
updates~\footnote{ Similar 
results can be obtained by starting from the fully ordered
state, $\sigma_i = 1$ for all $i$.}
[this because they equilibrate faster than (\ref{eq:10to01})].
Then, each protocol starts by switching instantaneously to $q/s^2=0.01$
at time $t=0$.
We set the parameter $B=1$, postponing the systematic study of different cases
to future works.
For the bubble exponent we use the value $c=2.14$~\cite{BCKMOS_pre03}.

\begin{figure}[!tb] 
\vskip 1truemm
\begin{center}
\includegraphics[angle=0,width=8.2cm]{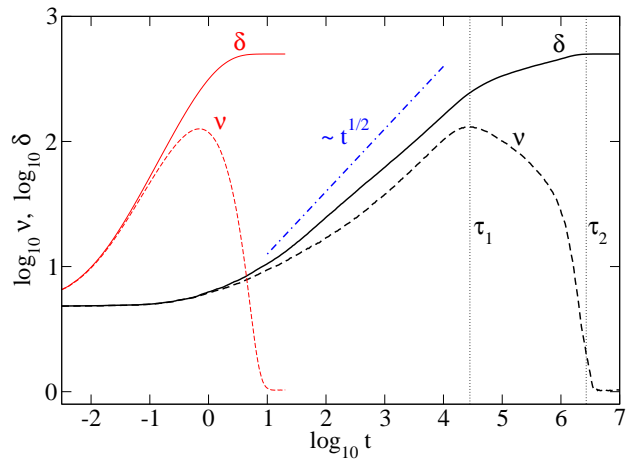}
\end{center}
\caption{(Color online)
Log-log plot of the number of open sites $\delta$ (dense lines) 
and of the number of bubbles $\nu$ (dashed lines) vs time, for $L=500$. 
Data shown with thick lines are obtained with our 
update~(\ref{eq:10to01})-(\ref{eq:sL}), 
while thin lines (red online) correspond to data obtained with 
update~(\ref{eq:1to0}). 
The final state for both dynamics is the equilibrium at high $T$,
with $\delta\approx L$  and $\nu\approx 1$.
The dot-dashed line represents a scaling  $\sim t^{1/2}$.
\label{fig:N500}} 
\end{figure}

The transient to the new
equilibrium is monitored by studying the scaling properties of 
two quantities, the number of denatured pairs 
$\delta\equiv \sum_i(1-\sigma_i)$ 
and the number of bubbles $\nu$. The former is 
the quantity normally inferred from UV-absorption
experiments~\cite{Wartell85} while $\nu$ is useful 
for characterizing the state of the system.
For convenience, data are binned in time intervals with 
constant size $=\log 1.05$ in log-scale. 
Moreover, an average over at least $1000$ trajectories is perfomed.
Fig.~\ref{fig:N500} shows $\delta$ and $\nu$ vs time 
in log-log scale for $L=500$,
both for a dynamics involving only move (\ref{eq:1to0}) and for 
the dynamics  (\ref{eq:10to01}) introduced in this paper.
In the former case, a fast denaturation takes place, at a
time scale $\tau_0\approx 1$ that does not scale with the
system size $L$~\cite{Kunz_LS_2007}.

The dynamics (\ref{eq:10to01})-(\ref{eq:sL}) generates a richer picture.
Two characteristic time scales have been highlighted by vertical lines
in Fig.~\ref{fig:N500}. 
The process goes as follows:
at high $T$, the rate $\sigma_1 = 1 \to \sigma_1 = 0$ is much higher then 
the rate of the opposite transition. The same is true for site $i=L$.
Thus,  $0$'s
enter at the boundaries and diffuse toward the center of the chain.
This first regime is characterized by an increase of $\delta$ and $\nu$
that is slower than $\sim t^{1/2}$~\footnote{In this regime we do not observe
simple dynamical scaling for $\delta$ and $\nu$.}, see Fig.~\ref{fig:N500}.
A time scale $\tau_1$ is characterized by the maximum of the number of
bubbles and it marks the end of the first regime.
At times  $t > \tau_1$ one observes
a second regime in which $\delta$ continues to increase while $\nu$ decreases,
which implies that bubbles coalesce. Finally, equilibrium is reached at a 
time $\tau_2$, when  $\delta \approx L$ and $\nu \approx 1$.

\begin{figure}[!tb] 
\vskip 1truemm
\begin{center}
\includegraphics[angle=0,width=8.9cm]{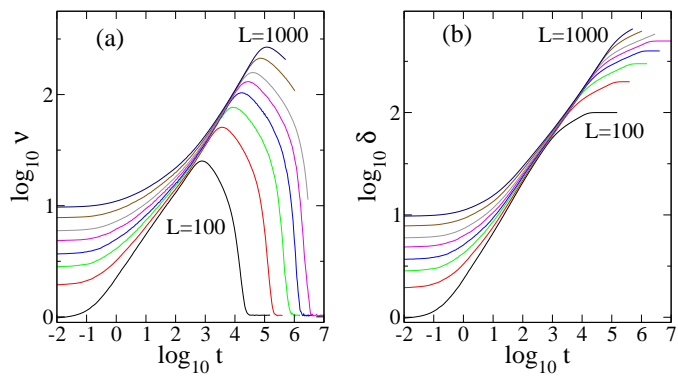}
\end{center}
\caption{(Color online)
(a) Number of bubbles as a function of time, for chain lengths
$L=100$, $200$, $300$, $400$, $500$, $600$, $800$ and $1000$ 
(from bottom to top). 
(b) Number of open sites vs $t$, same notation.
\label{fig:delta-nu}} 
\end{figure} 

\begin{figure}[!tb] 
\vskip 1truemm
\begin{center}
\includegraphics[angle=0,width=8.9cm]{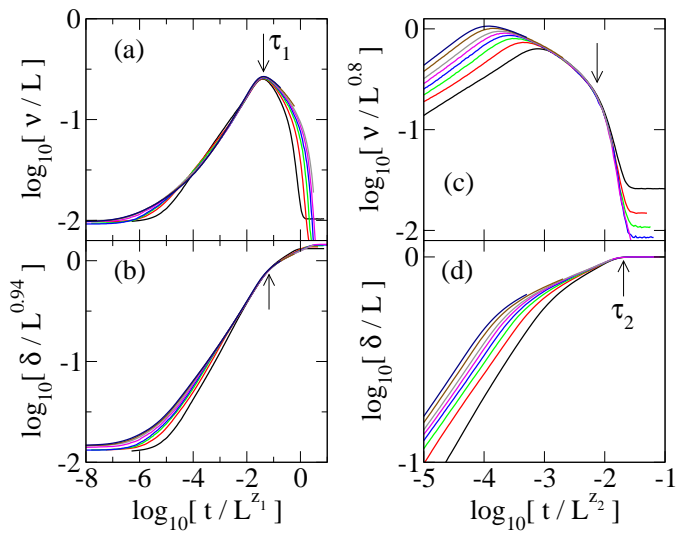}
\end{center}
\caption{(Color online)
Curves rescaled to collapse at $\tau_1$ with $z_1=2.15$,
and at $\tau_2$ with $z_2=3$,
see the notation of Fig.~\ref{fig:delta-nu} and 
Eqs.~(\ref{eq:z1})-(\ref{eq:z2}). 
\label{fig:resc12}} 
\end{figure}

Both $\tau_1$ and $\tau_2$ increase with the system size, see 
Fig.~\ref{fig:delta-nu}.
We find that both time scales are consistent with an algebraic dependence
on $L$, i.e.\
\begin{eqnarray}
\tau_1 \sim L^{z_1}\,,\qquad &\textrm{with }& z_1\simeq 2.15\pm 0.10 \label{eq:z1}\\
\tau_2 \sim L^{z_2}\,,\qquad &\textrm{with }& z_2\simeq 3.0\pm 0.1    \label{eq:z2}
\label{eq:tau}
\end{eqnarray}  
The peak of the $\nu$ plots 
is a particularly clear feature that helps to estimate the value
of $z_1$: peaks are reached at $\tau_1 \simeq 0.04 \times L^{2.15}$.
Moreover, around $\tau_1$ we achieve a data collapse of the form
$\nu / L$ vs $t / L^{z_1}$: Fig.~\ref{fig:resc12}(a) shows
that a critical density of bubbles $\nu(\tau_1) / L$
is reached at $\tau_1$. 
The values of the critical density of bubbles, as a function of
$L^{-1/2}$ and extrapolated for $L\to \infty$, tend to $0.280(1)$.
A similar collapse $\delta / L^{\alpha_1^\delta}$ vs $t / L^{z_1}$ can be attained:
the exponent yielding the best rescaling is $\alpha_1^\delta\simeq 0.94$, 
see Fig.~\ref{fig:resc12}(b).

Data collapses can be done also to determine $\tau_2$.
At  $\tau_2$ by definition we have the full denaturation,
i.e.~$\delta(\tau_2) \sim L$. 
To estimate $z_2\simeq 3$ we have in fact required that 
$\delta(t) / L \to 1$ for $t\to \tau_2$ (for all $L$'s for which $\tau_2$ 
could be reached by simulations), as shown in Fig.~\ref{fig:resc12}(d).
A data collapse of $\nu / L^{\alpha_2^\nu}$ vs $t/L^3$, with  $\alpha_2^\nu = 0.8$
[Fig.~\ref{fig:resc12}(c)], confirms that  $z_2 \simeq 3$ is the exponent 
characterizing the time scale $\tau_2$.

The first regime is essentially a diffusion of random walkers, the 
$\sigma_i = 0$ entering from the boundary, 
and one should expect a temporal domain scaling as the square 
of the system size. 
Indeed, we estimate $z_1=2$ for $c=0$, a case in which the open sites are 
independent of each other and the weight of a configuration
just depends on $\delta$.  
For $c=2.14$ we instead estimate the small deviation $z_1\simeq 2.15$
from this classical result, probably due to the bubble weights.
The case $c=0$ is also interesting because it does not display two 
different time scales but only one.
It confirms that the bubble interaction and coalescence is the process leading
to that second time scale. 

\begin{figure}[!tb] 
\vskip 1truemm
\begin{center}
\includegraphics[angle=0,width=8cm]{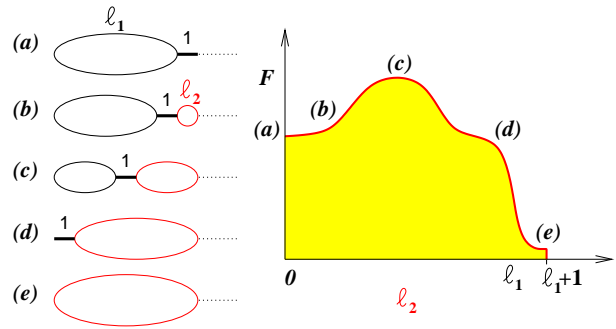}
\end{center}
\caption{(Color online)
Sketch of the exemplified process of escape of double-stranded segments.
Snapshots of some intermediate states and the corresponding 
free-energy profile (as a function of the length $\ell_2$ of the
growing loop) are shown. 
\label{fig:4}} 
\end{figure}

It is possible to predict the value of $z_2$ with
a simple argument: in the regime between $\tau_1$ and $\tau_2$ two bubbles
at the chain ends trap the double-stranded 
parts inside the system,  preventing a fast
escape of $\sigma_i=1$ from the boundaries. Let us concentrate on one of the
two ends, say $i=1$, as shown in Fig.~\ref{fig:4}, where an exemplified
escape of a double-stranded segment is shown. 
Taking the length $\ell_2$ of the forming loop on the right as a reaction 
coordinate, the free-energy $F = -\ln W$ has a
profile like the one shown in Fig~\ref{fig:4}.
It achieves a maximum at state (c), where $\ell_2=\ell_1/2$.
The rate of escape from (a) to (e) is proportional to the barrier jump rate,
which is proportional to the ratio of the weights in (c) and (a),
$[(\ell_1/2)^{-c}]^2 / \ell_1^{-c} \sim \ell_1^{-c}$. Hence the time
spent for this escape scales as $ (\ell_1)^c$.
As this has to take place for all $\ell_1$ up to the system size, 
$\tau_2 \sim \sum_{\ell_1\approx 1}^{L/2} (\ell_1)^c \sim L^{c+1}$, 
which would imply  $z_2 = c+1 \simeq 3.14$ 
if the whole dynamics was as simple as the one described.
Of course, multiple pathways intersect and a description only in terms of a
single $\ell_1$ might not be exhaustive. Nevertheless, 
this prediction is remarkably close to the value $z_2\simeq 3$ 
obtained by data collapse, suggesting that 
this is the main mechanism acting in the second regime.

Besides time scales, we have also a time dependence of the number of
open bases $\delta$ that is different from the one predicted by previous 
models, where one can observe a linear increase of 
$\delta$ with time~\cite{Marenduzzo_2002,Barbi_PRE2003} or 
$\delta\sim t^{3/4}$~\cite{Marenduzzo_2002},
in conditions similar to the one discussed in this
paper (initial state at low $T$, denaturation at high $T$, no external forces).
In our model, on the other hand, we observe that $\delta$ increases slower
than the square root of time.

The fact that polymer chains cannot cross each other is at the basis of
our version of the PS model, but of course it is included in many other 
models, like the model of polymer diffusing into a gel by 
de Gennes~\cite{deGennes89:_gel},
in which he found that the diffusion constant scales as $1/L^2$.
Furthermore, in simulations of polymers in dense 
melts~\cite{Paul91:_rept_simul} one observes autocorrelation times scaling
as $L^3$.
These long time scales derive from the reptation dynamics of the polymers,
which have to diffuse into tubes formed by the melt.
We argue that scenarios like that one are similar to the phenomenon 
predicted by our DNA model,
because the two twisted strands constrain the 
stochastic movements of each other in space during melting.
 
\subsection*{Acknowledgments}
We acknowledge useful discussions with E.~Orlandini, A.~Kabak\c c\i o\u glu,
P.~De Los Rios, A.~Flammini, F.~Piazza, C.~Maes, E.~Carlon, and G.~T.~Barkema.
We also thank E. Carlon for his useful comments on the manuscript.
M.B. acknowledges 
support from EC FP6 project 
``EMBIO'' (EC contract nr.~012835) and from K.U.Leuven grant OT/07/034A.


\begin{thebibliography}{10}

\bibitem{Alberts02:_book}
B.~Alberts et~al.
\newblock {\em Molecular Biology of the Cell}.
\newblock Gerland Science, New York, 2002.

\bibitem{Wartell85}
R.~M. Wartell and A.~S. Benight.
\newblock {\em Phys.\ Rep.}, 85:67, 1985.

\bibitem{PS_JCP1966}
D.~Poland and H.~Scheraga.
\newblock {\em J. Chem. Phys.}, 45:1464, 1966.

\bibitem{Blake99:_meltsim}
R.~D. Blake et~al.
\newblock {\em Bioinformatics}, 15:370, 1999.

\bibitem{PB_PRL1989}
M.~Peyrard and A.~R. Bishop.
\newblock {\em Phys. Rev. Lett.}, 62:2755, 1989.

\bibitem{Barbi_PRE2003}
M.~Barbi, S.~Lepri, M.~Peyrard, and N.~Theodorakopoulos.
\newblock Thermal denaturation of a helicoidal dna model.
\newblock {\em Phys. Rev. E}, 68:061909, 2003.

\bibitem{Drukker_WS_JCP2001}
K.~Drukker, G.~Wu, and G.~C. Schatz.
\newblock Model simulations of {DNA} denaturation dynamics.
\newblock {\em J. Chem. Phys.}, 114:579--590, 2001.

\bibitem{Knotts_JCP2007}
T.~A. {Knotts IV}, N.~Rathore, D.~C. Schwartz, and J.~J. {de Pablo}.
\newblock A coarse grain model of {DNA}.
\newblock {\em J. Chem. Phys.}, 126:084901, 2007.

\bibitem{RudnickBruinsma_PRE2002}
J.~Rudnick and R.~Bruinsma.
\newblock Effects of torsional strain on thermail denaturation of {DNA}.
\newblock {\em Phys. Rev. E}, 65:030902(R), 2002.

\bibitem{alkan2008}
A.~{Kabak\c{c}{\i}o\u{g}lu}, E.~Orlandini, and D.~Mukamel.
\newblock Supercoil formation in {DNA} denaturation, 2008.
\newblock arxiv:0811.3229.

\bibitem{Heukelum02:_electroph}
A.~van Heukelum and G.~T. Barkema.
\newblock Lattice models of dna electrophoresis.
\newblock {\em Electrophor.}, 23:2562--2568, 2002.

\bibitem{Rubinstein87}
M.~Rubinstein.
\newblock {\em Phys. Rev. Lett.}, 59:1946--1949, 1987.

\bibitem{Duke89}
T.~A.~J. Duke.
\newblock {\em Phys. Rev. Lett.}, 62:2877--2880, 1989.

\bibitem{Carlon01:_RDmodel}
E.~Carlon, A.~Drzewinski, and J.~M.~J. {van Leeuwen}.
\newblock Crossover behavior for long reptating polymers.
\newblock {\em Phys. Rev. E}, 64:010801, 2001.

\bibitem{Marenduzzo_2002}
D.~Marenduzzo, S.~M. Bhattacharjee, A.~Maritan, E.~Orlandini, and F.~Seno.
\newblock Dynamical scaling of the dna unzipping transition.
\newblock {\em Phys. Rev. Lett.}, 88(2):028102, 2002.

\bibitem{Kunz_LS_2007}
H.~Kunz, R.~Livi, and A.~{S\"uto}.
\newblock The structure factor and dynamics of the helix-coil transition.
\newblock {\em J. Stat. Mech.}, 2007(06):P06004, 2007.

\bibitem{Hanke03:_bubble_dyn}
Andreas Hanke and Ralf Metzler.
\newblock Bubble dynamics in dna.
\newblock {\em J. Phys. A: Math. Gen.}, 36:L473, 2003.

\bibitem{Novotny06:_vicious}
Tomas Novotny, Jonas~Nyvold Pedersen, Tobias Ambjornsson, Mikael~Sonne Hansen,
  and Ralf Metzler.
\newblock Bubble coalescence in breathing dna: Two vicious walkers in opposite
  potentials.
\newblock {\em Europhys. Lett.}, 77:48001, 2007.

\bibitem{Altan-bonnet03}
G.~{Altan-Bonnet}, A.~Libchaber, and O.~Krichevsky.
\newblock {\em Phys. Rev. Lett.}, 90:138101, 2003.

\bibitem{Kafri_MP_2000}
Y.~Kafri, D.~Mukamel, and L.~Peliti.
\newblock {\em Phys. Rev. Lett.}, 85:4988, 2000.

\bibitem{Dean85:_topois}
F.~B. Dean, A.~Stasiak, T.~Koller, and N.~R. Cozzarelli.
\newblock {\em J. Biol. Chem.}, 260:4975, 1985.

\bibitem{BCKMOS_pre03}
M.~Baiesi, E.~Carlon, Y.~Kafri, D.~Mukamel, E.~Orlandini, and A.~L. Stella.
\newblock Inter-strand distance distribution of {DNA} near melting.
\newblock {\em Phys. Rev. E}, 67:021911, 2003.

\bibitem{deGennes89:_gel}
P.~G. {de Gennes}.
\newblock {\em J. Chem. Phys.}, 91:3252--3257, 1989.

\bibitem{Paul91:_rept_simul}
W.~Paul, K.~Binder, D.~W. Heermann, and K.~Kremer.
\newblock {\em J. Chem. Phys.}, 95:7726--7740, 1991.

\end{thebibliography}

\end{document}